\documentclass[12pt]{article}

\usepackage[dvips]{graphicx}
\usepackage{latexsym}

\usepackage{amssymb,amsfonts,amsmath}
\usepackage{graphicx} 
\usepackage{indentfirst}

 \usepackage{bbm}

\parskip=\medskipamount

\newcommand {\cA}{{\cal A}}

\newcommand {\cD}{{\cal D}}

\newcommand {\cF}{{\cal F}}

\newcommand {\cH}{{\cal H}}

\newcommand {\cL}{{\cal L}}
\newcommand {\cM}{{\cal M}}
\newcommand {\cN}{{\cal N}}

\newcommand {\cQ}{{\cal Q}}
\newcommand {\cR}{{\cal R}}

\newcommand {\cT}{{\cal T}}

\newcommand {\cV}{{\cal V}}

\def\a{\alpha}
\def \bi{\bibitem}

\def\b{\beta}

\def\d{\delta}
\def\e{\epsilon}
\def\f{\phi}
\def\g{\gamma}
\def\G{\Gamma}

\def\l{\lambda}
\def\m{\mu}
\def\n{\nu}
\def\o{\omega}
\def\p{\pi}
\def\q{\theta}
\def\r{\rho}
\def\s{\sigma}
\def\t{\tau}

\def\z{\zeta}
\def\D{\Delta}
\def\F{\Phi}
\def\J{\Psi}
\def\L{\Lambda}
\def\O{\Omega}

\def\S{\Sigma}
\def\U{\Upsilon}
\def\X{\Xi}

\def\rd{{\rm d}}
\def\ri{{\rm i}}

\newcommand{\ve}{\varepsilon}                            

\newcommand{\pa}{\partial}                           
\newcommand{\hf}{\frac12}

\newcommand{\sect}[1]{\setcounter{equation}{0}\section{#1}}

\newcommand{\be}{\begin{equation}}
\newcommand{\ee}{\end{equation}}
\newcommand{\bea}{\begin{eqnarray}}
\newcommand{\eea}{\end{eqnarray}}
\newcommand{\non}{\nonumber}
\newcommand{\1}{\underline{1}}
\newcommand{\2}{\underline{2}}

\newcommand{\bm}[1]{\mbox{\boldmath$#1$}}

\def\double #1{#1{\hbox{\kern-2pt $#1$}}}

\newcommand{\hm}{{\hat{m}}}
\newcommand{\hn}{{\hat{n}}}

\newcommand{\ha}{{\hat{a}}}
\newcommand{\hb}{{\hat{b}}}
\newcommand{\hc}{{\hat{c}}}
\newcommand{\hd}{{\hat{d}}}
\newcommand{\he}{{\hat{e}}}

\newcommand{\hal}{{\hat{\a}}}
\newcommand{\hbe}{{\hat{\b}}}
\newcommand{\hga}{{\hat{\g}}}
\newcommand{\hde}{{\hat{\d}}}
\newcommand{\hrh}{{\hat{\rho}}}

\newcommand{\CD}{{\nabla}}

\begin{document}

\begin{titlepage}

\begin{flushright}
October, 2007\\
Revised version: January, 2008\\
\end{flushright}
\vspace{5mm}

\begin{center}
{\Large \bf  Five-dimensional 
Superfield 
Supergravity }\\ 
\end{center}

\begin{center}

{\large  
Sergei M. Kuzenko\footnote{{kuzenko@cyllene.uwa.edu.au}}
and 
Gabriele Tartaglino-Mazzucchelli\footnote{gtm@cyllene.uwa.edu.au}
} \\
\vspace{5mm}

\footnotesize{
{\it School of Physics M013, The University of Western Australia\\
35 Stirling Highway, Crawley W.A. 6009, Australia}}  
~\\

\vspace{2mm}

\end{center}
\vspace{5mm}

\begin{abstract}
\baselineskip=14pt
We present a projective superspace formulation for 
matter-coupled simple supergravity in five dimensions.
Our starting  point is the superspace realization for the  minimal 
supergravity multiplet proposed by Howe in 1981.
We introduce various off-shell supermultiplets 
(i.e. hypermultiplets, tensor and vector multiplets)
that describe matter fields coupled to supergravity.
A projective-invariant action principle is given, and specific 
dynamical systems are constructed including supersymmetric 
nonlinear sigma-models. We believe that this approach can 
be extended to  other supergravity theories with eight supercharges
in $D\leq 6$ space-time dimensions, including 
the important case of 4D $\cN=2$ supergravity. 
\end{abstract}
\vspace{1cm}

\vfill
\end{titlepage}

\newpage
\renewcommand{\thefootnote}{\arabic{footnote}}
\setcounter{footnote}{0}


\sect{Introduction}
Projective superspace \cite{KLR,LR1} is a powerful formalism 
for building off-shell rigid supersymmetric theories with eight supercharges
in $D\leq 6$ space-time dimensions. 
It is ideal  for the  
explicit construction of hyperk\"ahler metrics \cite{HitKLR}.
${}$For more than two decades, it has been
an open problem to extend this approach to supergravity. 
A partial success has been achieved in our recent paper \cite{KT-M} 
where, in particular,  the relevant  projective formulation was developed 
for  supersymmetric theories in 
five-dimensional $\cN=1$ anti-de Sitter superspace
AdS${}^{5|8}={\rm SU}$(2,2$|$1)/SO$(4,1)\times {\rm U}(1)$ which is  a maximally symmetric
curved background. 
In this letter we briefly describe a solution to the problem 
in the case of 5D simple supergravity. A more detailed presentation 
will be given elsewhere \cite{KT-M2}.

${}$For 5D $\cN=1$ supergravity\footnote{On historical grounds, 
5D simple ($\cN=1$) supersymmetry and supergravity are often labeled $\cN=2$.} 
\cite{Cremmer}, off-shell superspace formulations were  only sketched by 
Breitenlohner and Kabelschacht \cite{Breitenlohner}  
and  independently by Howe  \cite{Howe5Dsugra} 
(who built on the 5D supercurrent constructed in  \cite{HL}).
Later, general matter couplings in  5D simple supergravity 
were constructed within on-shell components approaches \cite{GST,GZ,CD}.
More recently, off-shell component formulations for 5D supergravity-matter systems
were developed in \cite{Zucker} and independently, 
within the superconformal tensor calculus, in \cite{Ohashi,Bergshoeff}.
Since the approaches elaborated in \cite{Zucker,Ohashi,Bergshoeff} are intrinsically 
component (i.e. they make use of 
off-shell hypermultiplets with finitely many 
auxiliary fields and an intrinsic central 
charge), they do not allow us to construct the most general sigma-model couplings, 
similar to the four-dimensional  $\cN=2$ case, and thus a superspace 
description is still desirable. Such a formulation is given below.

Before turning to the description of our superspace approach, we should emphasize 
once more 
that  it is the presence of  the intrinsic central charge that hypermultilets possess, 
within the component formulations of   \cite{Zucker,Ohashi} which makes it impossible
to cast general quaternionic K\"ahler couplings in terms of such
off-shell hypermultiplets\footnote{Refs. \cite{Bergshoeff} 
deal with on-shell hypermultiplets only.} 
(see, e.g,,  \cite{Galicki} for a similar discussion in the case of 4D $\cN=2$ supergravity).
On the other hand, the projective superspace approach offers
nice off-shell formulations without central charge. 
Specifically, there are infinitely many  off-shell realizations with {\it finitely many auxiliary fields}
 for a neutral hypermultiplet (they are the called $O(2n)$ multiplets,  where $n=2, 3\dots$, 
 following the terminology 
of \cite{G-RLRvUW}), and a unique formulation for a charged hypermultiplet 
with {\it infinitely many auxiliary fields}
(the so-called polar hypermultiplet).
Using covariant polar hypermultiplets introduced below, 
one can construct  sigma-model couplings 
that cannot be derived within the off-shell component approach. 
An example is given by the supergravity-matter system (\ref{SUGRA-matter}),
for a generic choice of the real function $K(\F, \bar \F) $ 
obeying the homogeneity condition (\ref{Kkahler2}).
We hope to discuss this point in more detail elsewhere.

This paper is organized as follows. 
In section 2,  we describe the superspace geometry of the minimal multiplet 
for 5D $\cN=1$ supergravity outlined in  \cite{Howe5Dsugra}. 
Various off-shell supermultiplets 
are introduced in section 3. In section 4, we present a supersymmetric  
projective-invariant action principle in a Wess-Zumino gauge. 
This action is ready for applications, in particular
for a reduction from superfields to components. 
We also introduce several families of supergravity-matter systems. Finally, in section 5
we describe a locally supersymmetric action which we expect  
to reduce, in the Wess-Zumino gauge, 
to the action given in section 4.

\sect{Superspace geometry of the minimal supergravity multiplet}

Let $z^{\hat{M}}=(x^{\hm},\q^{\hat{\mu}}_i)$
be local bosonic ($x$) and fermionic ($\q$) 
coordinates parametrizing  a curved five-dimensional $\cN=1$  superspace
$\cM^{5|8}$,
where $\hm=0,1,\cdots,4$, $\hat{\mu}=1,\cdots,4$, and  $i=\1,\2$.
The Grassmann variables $\q^{\hat{\mu}}_i$
are assumed to obey a standard pseudo-Majorana reality condition. 
${}$Following \cite{Howe5Dsugra}, the tangent-space group
is chosen to be  ${\rm SO}(4,1)\times {\rm SU}(2)$,
and the superspace  covariant derivatives 
$\cD_{\hat{A}} =(\cD_{\hat{a}}, \cD_{\hat{\a}}^i)$
have the form 
\bea
\cD_{\hat{A}}&=&E_{\hat{A}} + \O_{\hat{A}} + \F_{\hat{A}}
+V_{\hat{A}}Z~.
\label{CovDev}
\eea
Here $E_{\hat{A}}= E_{\hat{A}}{}^{\hat{M}}(z) \,\pa_{\hat{M}}$ is the supervielbein, 
with $\pa_{\hat{M}}= \pa/ \pa z^{\hat{M}}$,
\bea
\O_{\hat{A} }= \hf \,\O_{\hat{A}}{}^{\hb\hc}\,M_{\hb\hc}
= \O_{\hat{A}}{}^{\hbe\hga}\,M_{\hbe\hga}~,\qquad 
M_{\ha\hb}=-M_{\hb\ha}~, \quad M_{\hal\hbe}=M_{\hbe\hal}
\eea
is the Lorentz connection, 
\bea
\F_{\hat{A}} = \F^{~\,kl}_{\hat{A}}\,J_{kl}~, \qquad
J_{kl}=J_{lk}
\eea
is the SU(2)-connection, and $Z$ the central-charge generator.
The Lorentz generators with vector indices ($M_{\ha\hb}$) and spinor indices
($M_{\hal\hbe}$) are related to each other by the rule:
$M_{\ha\hb}=(\S_{\ha\hb})^{\hal\hbe}M_{\hal\hbe}$ 
(for more details regarding our 5D notation and conventions, see the appendix in \cite{KT-M}).
The generators of ${\rm SO}(4,1)\times {\rm SU}(2)$
act on the covariant derivatives as follows:
\bea
{[}J^{kl},\cD_{\hal}^i{]}
= \ve^{i(k} \cD^{l)}_{\hat \a}~,
\qquad 
{[}M_{\hal\hbe},\cD_{\hga}^i{]}
=\ve_{\hga(\hal}\cD^i_{\hbe)}~,
\eea
where $J^{kl} =\ve^{ki}\ve^{lj} J_{ij}$,  and the symmetrization of $n$ indices
involves a factor of $(n!)^{-1}$. 
The covariant derivatives obey  (anti)commutation relations of the general form 
\bea
{[}\cD_{\hat{A}},\cD_{\hat{B}}\}&=&T_{\hat{A}\hat{B}}{}^{\hat{C}}\cD_{\hat{C}}
+\hf R_{\hat{A}\hat{B}}{}^{\hat{c}\hat{d}}M_{\hat{c}\hat{d}}
+R_{\hat{A}\hat{B}}{}^{kl}J_{kl}
+F_{\hat{A}\hat{B}}Z~,
\label{algebra}
\eea
where $T_{\hat{A}\hat{B}}{}^{\hat{C}}$ is the torsion, 
$R_{\hat{A}\hat{B}}{}^{kl}$ and $R_{\hat{A}\hat{B}}{}^{\hat{c}\hat{d}}$  
the SU(2)- and SO(4,1)-curvature tensors, 
and $F_{\hat{A}\hat{B}}$  the central charge field strength.

The supergravity gauge group is generated by local transformations
of the form 
\be
\cD_{\hat{A}} \to \cD'_{\hat{A}}  ={\rm e}^{K  }\, \cD_{\hat{A}}\, {\rm e}^{-K  } ~,
\qquad K = K^{\hat{C}}(z) \cD_{\hat{C}} +\hf K^{\hat c \hat d}(z) M_{\hat c \hat d}
+K^{kl}(z) J_{kl}  +\t(z) Z~,
\label{tau}
\ee
with all the gauge parameters being neutral with respect to the central charge $Z$,
obeying natural reality conditions, and otherwise  arbitrary. 
Given a tensor superfield $U(z)$, it transforms as follows:
\bea
U \to U' = {\rm e}^{K  }\, U~.
\eea

In accordance with \cite{Howe5Dsugra},  in order to to realize the so-called minimal supergravity 
multiplet in the above framework, one has to impose special covariant  constraints on 
various components of the torsion\footnote{As demonstrated by
Dragon  \cite{Dragon}, the curvature is completely determined in terms 
of the torsion in supergravity 
theories formulated in superspace.}  of dimensions 0, 1/2 and 1.
They are:
\begin{subequations}
\bea
T_{\hat \a}^i{}_{\hat \b}^j\,{}^{\hc}=-2\ri\, \ve^{ij}(\Gamma^{\hc})_{\hal\hbe}~, \qquad
F_{\hat \a}^i\,{}_{\hat  \b}^j=-2\ri\,
\ve^{ij}\ve_{\hal\hbe}~,  \qquad && \mbox{(dimension-0)}  \\
T_{\hat \a}^i\,{}_{\hat \b}^j\,{}^{\hat \g}_k=
T_{\hat \a}^i\,{}_{\hb}{}^{\hc}=
F_{\hat \a}^i\,{}_{\hb}=0~,  \quad \qquad \qquad && \mbox{(dimension-1/2)}  \\
T_{\ha\hb}\,{}^{\hc}~=~T_{\ha\hbe (j}{}^{\hbe}{}_{k)}~=~0~. \quad \qquad 
\qquad && \mbox{(dimension-1)}
\eea
\end{subequations}
Under these constraints, the algebra (\ref{algebra}) can be shown to 
take the form\footnote{In \cite{Howe5Dsugra}, the  solution 
to the constraints was not given in detail. 
In particular, the algebra of covariant derivatives
(\ref{covDev2spinor-}--\ref{covDev2spinor-3}) was not included.   }
(its derivation will be given in \cite{KT-M2}):
\begin{subequations}
\bea
\big\{ \cD_{\hal}^i , \cD_{\hbe}^j \big\} &=&-2 \ri \,\ve^{ij}\cD_{\hal\hbe}
-2\ri \, \ve^{ij}\ve_{\hal\hbe}Z
\non\\
&&
+3\ri \, \ve_{\hal\hbe}\ve^{ij}S^{kl}J_{kl}
-2\ri(\S^{\ha\hb})_{\hal\hbe}
\Big(F_{\ha\hb}+N_{\ha\hb}\Big)J^{ij}
\non\\
&&
-\ri \,\ve_{\hal\hbe}\ve^{ij}F^{\hc\hd}M_{\hc\hd}
+{\ri\over 4} \ve^{ij}\ve^{\ha\hb\hc\hd\he}N_{\ha\hb}(\G_\hc)_{\hal\hbe}M_{\hd\he}
+4\ri \,S^{ij}M_{\hal\hbe}
~,
\label{covDev2spinor-} \\
{[}\cD_\ha,\cD_{\hbe}^j{]}&=&
{1\over 2}(\Gamma_{\hat{a}})_{\hbe}{}^{\hga}S^j{}_k\cD_{\hga}^k
-{1\over 2}\,F_{\ha\hb}(\Gamma^{\hat{b}})_{\hbe}{}^{\hga}\cD_{\hga}^j
-{1\over 8}\,\ve_{\ha\hb\hc\hd\he}N^{\hd\he}(\Sigma^{\hb\hc})_{\hbe}{}^{\hga}\cD_{\hga}^j
\non\\
&&
+\Big(-3\ve^{jk}\X_{\ha\hbe}{}^{l}
+{5\over 4} (\G_{\ha})_{\hbe}{}^{\hal}\ve^{jk}\cF_{\hal}{}^{l}
-{1\over 4}(\G_\ha)_{\hbe}{}^{\hal}\ve^{jk}\cN_\hal{}^{l}\Big) J_{kl}
\non\\
&&
+\Big(
{1\over 2}(\G_{\ha})^{\hal\hga}\cD^{\hde j} F_{\hal\hbe}
-{1\over 2}(\G_{\ha})_{\hbe\hal}\cD^{\hga j} F^{\hde\hal}
-{1\over 2}(\G_{\ha})^{\hal\hga}\d^{\hde}_{\hbe}\cD^{\hrh j} F_{\hal\hrh}
\non\\
&&
+{1\over 2}(\G_{\ha})_{\hrh\hal}\cD^{\hrh j} F^{\hal\hga}\d^{\hde}_{\hbe}
-{1\over 2} (\G_\ha)_{\hbe\hrh}\cD^{\hrh j}F^{\hga\hde}
\Big)
M_{\hga\hde}~, 
\label{covDev2spinor-2}\\
{[}\cD_\ha,\cD_\hb{]}&=&
{\ri\over 2} \Big(\cD^\hga_kF_{\ha\hb}\Big)\cD_\hga^k 
-{\ri\over 8}\Big(\cD^{\hga (k} \cD_\hga^{l)}F_{\ha\hb}\Big)J_{kl}
+F_{\ha\hb}Z
\non\\
&&
+\Big(\,{1\over 4}\ve^{\hc\hd}{}_{\hm\hn{[}\ha}\cD_{\hb{]}}N^{\hm\hn}
+\hf\d^{\hc}_{{[}\ha}N_{\hb{]}\hm}N^{\hd\hm}
-{1\over 4}N_\ha{}^\hc N_\hb{}^{\hd}
-{1\over 8}\d_\ha^{\hc}\d^{\hd}_{\hb}N^{\hm\hn}N_{\hm\hn}
\non\\
&&~~~
+{\ri\over 8}(\S^{\hc\hd})^{\hga\hde}\cD_{\hga}^k \cD_{\hde k}F_{\ha\hb}
-F_{\ha}{}^{\hc}F_{\hb}{}^{\hd}
+{1\over 2}\d_\ha^{\hc}\d^{\hd}_{\hb}S^{ij}S_{ij}\Big)
M_{\hc\hd}~.
\label{covDev2spinor-3}
\eea
\end{subequations}
The torsion components obey a number of Bianchi idenities some of which 
can be conveniently  expressed in terms of 
the three irreducible components of  $\cD_\hga^kF_{\hal\hbe}$: a completely 
symmetric third-rank tensor $W_{\hal\hbe\hga}{}^k$,
a gamma-traceless
spin-vector $\X_{ \ha \,\hga}{}^k$ and a spinor $\cF_{\hga}{}^k$.
These components originate as follows:
\bea
\cD_\hga^kF_{\hal\hbe}&=&
W_{\hal\hbe\hga}{}^k
+(\G_\ha)_{\hga(\hal}\X^\ha{}_{\hbe)}{}^k
+\ve_{\hga(\hal}\cF_{\hbe)}{}^k~,\non \\
&&W_{\hal\hbe\hga}{}^k=W_{( \hal\hbe\hga )}{}^k~, \qquad 
(\G^\ha)_\hal{}^\hbe\X_{\ha\hbe}{}^i=0~.
\eea
The dimension-3/2  Bianchi identities 
are as folllows:
\begin{subequations}
\bea
\cD_\hga^kN_{\hal\hbe}&=&
-W_{\hal\hbe\hga}{}^k
+2(\G_\ha)_{\hga(\hal}\X^\ha{}_{\hbe)}{}^k
+\ve_{\hga(\hal}\cN_{\hbe)}{}^k~,
\label{N-or-Bianchi}\\
\cD_{\hbe}^{k}S^{jl}&=&
-\hf \ve^{k(j}\Big(3\cF_\hbe{}^{l)}+\cN_{\hbe}{}^{l)}\Big)~.
\label{S-Bianchi} 
\eea
\end{subequations}
The   Bianchi identities 
of dimension 2 will be described in \cite{KT-M2}.
A simple consequence of (\ref{S-Bianchi}) is 
\bea
\cD_{\hal}^{(i}S^{jk)} =0~.
\label{S-Bianchi2}
\eea
This result will be important  in what follows.

\sect{Projective supermultiplets}

To introduce  an important class of off-shell supermultiplets,
it is convenient to  make use of  an  isotwistor  $u^{+}_i \in  
{\mathbb C}^2 \setminus  \{0\}$ defined to be inert with respect to 
the local SU(2) group (in complete analogy with \cite{KT-M,K2}).
Then, in accordance with (\ref{covDev2spinor-}), the operators
$\cD^+_{\hat \a}:=u^+_i\,\cD^i_{\hat \a} $  obey the following algebra:
\bea
\{  \cD^+_{\hat \a} , \cD^+_{\hat \b} \}
=-4{\rm i}\, \Big(F_{\hal \hbe}+N_{\hal \hbe}\Big)\,J^{++}
+4{\rm i} \, S^{++}M_{\hal \hbe}~,
\label{analyt}
\eea
where 
$J^{++}:=u^+_i u^+_j \,J^{ij}$ and 
$S^{++}:=u^+_i u^+_j \,S^{ij}$.
Relation (\ref{analyt}) naturally hints at the possibility 
of introducing covariant superfields $Q(z,u^+)$ obeying 
the chiral-like condition $\cD^+_{\hat \a} Q =0$ (which is a generalization 
of the so-called analyticity condition in 4D $\cN=2$ rigid supersymmetric
\cite{GIKOS,KLR} and 5D $\cN=1$ anti-de Sitter \cite{KT-M}
cases). For this constraint to be consistent, however,
such superfields must be scalar with respect to the Lorentz group,
$ M_{\hal \hbe} Q=0$, and also possess special properties with respect to the
group SU(2), that is,  $J^{++}Q=0$. 
Now we define such multiplets.

A {\it covariantly analytic} multiplet of weight $n$,
$Q^{(n)}(z,u^+)$, is a scalar superfield that 
lives on  $\cM^{5|8}$, 
is holomorphic with respect to 
the isotwistor variables $u^{+}_i $ on an open domain of 
${\mathbb C}^2 \setminus  \{0\}$, 
and is characterized by the following conditions:\\
(i) it obeys the analyticity constraint 
\be
\cD^+_{\hat \a} Q^{(n)}  =0;
\label{ana}
\ee  
(ii) it is  a homogeneous function of $u^+$ 
of degree $n$, that is,  
\be
Q^{(n)}(z,c\,u^+)\,=\,c^n\,Q^{(n)}(z,u^+)~, \qquad c\in \mathbb{C}^*~;
\label{weight}
\ee
(iii) infinitesimal gauge transformations (\ref{tau}) act on $Q^{(n)}$ 
as follows:
\bea
\d Q^{(n)} 
&=& \Big( K^{\hat{C}} \cD_{\hat{C}} + K^{ij} J_{ij} 
+\t Z \Big)Q^{(n)} ~,  
\non \\ 
K^{ij} J_{ij}  Q^{(n)}&=& -\frac{1}{(u^+u^-)} \Big(K^{++} D^{--} 
-n \, K^{+-}\Big) Q^{(n)} ~, \qquad 
K^{\pm \pm } =K^{ij}\, u^{\pm}_i u^{\pm}_j ~,
\label{harmult1}   
\eea 
where
\bea
D^{--}=u^{-i}\frac{\partial}{\partial u^{+ i}} ~,\qquad
D^{++}=u^{+ i}\frac{\partial}{\partial u^{- i}} ~.
\label{5}
\eea
Transformation law (\ref{harmult1}) involves an additional isotwistor,  $u^-_i$, 
which is subject 
to the only condition $(u^+u^-) = u^{+i}u^-_i \neq 0$, and is otherwise completely arbitrary.
By construction, $Q^{(n)}$ is independent of $u^-$, 
i.e. $\pa  Q^{(n)} / \pa u^{-i} =0$,
and hence $D^{++}Q^{(n)}=0$.
It is easy to see that $\d Q^{(n)} $ 
is also independent of the isotwistor $u^-$, $\pa (\d Q^{(n)})/\pa u^{-i} =0$,
as a consequence of (\ref{weight}).
It follows from (\ref{harmult1}) 
\bea
J^{++} \,Q^{(n)}=0~, \qquad J^{++} \propto D^{++}~,
\label{J++}
\eea
and therefore the constraint (\ref{ana}) is indeed consistent.
It is important to point out  that eq. (\ref{J++}) is purely algebraic. 

In what follows, our consideration will be restricted to those 
supermultiplets that are inert with respect to the central charge, 
$ZQ^{(n)}=0$.

Given a covariantly analytic   superfield $Q^{(n)}$,
its complex conjugate 
is not analytic.
However, similarly to the flat four-dimensional case  
\cite{GIKOS,KLR}  (see also \cite{KT-M}),
one can introduce a generalized,  analyticity-preserving 
conjugation, $Q^{(n)} \to \widetilde{Q}^{(n)}$, defined as
\be
\widetilde{Q}^{(n)} (u^+)\equiv \bar{Q}^{(n)}\big(\widetilde{u}^+\big)~, 
\qquad \widetilde{u}^+ = {\rm i}\, \s_2\, u^+~, 
\ee
with $\bar{Q}^{(n)}$ the complex conjugate of $Q^{(n)}$.
Its fundamental property is
\bea
\widetilde{ {\cD^+_{\hat \a} Q^{(n)}} }=(-1)^{\e(Q^{(n)})}\, \cD^{+\hat \a}
 \widetilde{Q}{}^{(n)}~. 
\eea
One can see that
$\widetilde{\widetilde{Q}}{}^{(n)}=(-1)^nQ^{(n)}$,
and therefore real supermultiplets can be consistently defined when 
$n$ is even.
In what follows, $\widetilde{Q}^{(n)}$ will be called the smile-conjugate of 
${Q}^{(n)}$.

With respect to the natural projection
$\p\!\!: {\mathbb C}^2 \setminus  \{0\} \to  {\mathbb C}P^1$, 
the isotwistor $u^+_i$  plays the role of homogeneous global coordinates
for $ {\mathbb C}P^1$, and the covariantly analytic superfields
$Q^{(n)}(z,u^+)$ introduced describe special supermultiplets living in  
$\cM^{5|8} \times {\mathbb C}P^1$.
As is well-known, instead of the homogeneous coordinates $u^+_i$, it is often useful 
to work with an inhomogeneous local complex variable $\z$ that is invariant 
under arbitrary  projective rescalings  $u^+_i \to c\, u^+_i $, with $ c\in \mathbb{C}^*$.
In such an approach, one should replace $Q^{(n)}(z,u^+)$ with a new superfield 
$Q^{[n]}(z,\z) \propto Q^{(n)}(z,u^+)$, where $Q^{[n]}(z,\z) $ is  holomorphic 
with respect to  $\z$, and its explicit definition depends on the supermultiplet under 
consideration.  
The space ${\mathbb C}P^1$ can 
naturally be covered by two open charts in which $\z$ can be defined, 
and the simplest choice is:
(i) the north chart characterized by $u^{+\1}\neq 0$;
(ii) the south chart with  $u^{+\2}\neq 0$.
In discussing various supermultiplets, 
our consideration below will be restricted to the north chart.
 
In the north chart $u^{+\1}\neq 0$, 
and the projective-invariant  variable $\z \in \mathbb C$
can be defined in the simplest way:
\bea
u^{+i} =u^{+\1}(1,\z) =u^{+\1}\z^i ~,\qquad 
\z^i=(1,\z)~, \qquad \z_i= \ve_{ij} \,\z^j=(-\z,1)~.
\label{zeta}
\eea
Since any projective multiplet $Q^{(n)}$ and its
transformation (\ref{harmult1}) do not depend on $u^-$, 
we can make a convenient choice for the latter.
In the north chart, it is
\be
u^-_i =(1,0) ~, \qquad   \quad ~u^{-i}=\ve^{ij }\,u^-_j=(0,-1)~.
\label{fix-u-}
\ee   
The transformation parameters $K^{++}$  
and $K^{+-}$ in (\ref{harmult1}) 
can be represented as $K^{++} =\big(u^{+\1}\big)^2 {K}^{++} (\z)$ 
and $K^{+-}= u^{+\1}K(\z) $, where
\bea
{K}^{++} (\z)&=& {K}^{ij} \,\z_i \z_j
=  {K}^{\1 \1 }\, \z^2 -2  {K}^{\1 \2}\, \z 
+ {K}^{\2 \2} ~,
\quad
K(\z)= {K}^{\1 i} \,\z_i 
=  - {K}^{\1 \1} \,\z + {K}^{\1 \2}  ~.~~~~~~~
\label{K++K}
\eea
If the projective supermultiplet  $Q^{(n)}(z,u^+)$ 
is described by $Q^{[n]}(z,\z) \propto Q^{(n)}(z,u^+)$ in the north chart, 
then the analyticity condition (\ref{ana}) turns into 
\bea
\cD^+_{\hat \a} (\z) \, Q^{[n]} (\z) =0~, 
\qquad \cD^+_{\hat \a} (\z) = \cD^i_{\hat \a} \z_i
=-\z \,\cD^{\1}_{\hat \a}  + \cD^{\2}_{\hat \a} ~.
\label{ana2}
\eea
Let us give several important examples of projective supermultiplets.

An arctic multiplet\footnote{For covariantly analytic multiplets, 
we adopt  the same terminology 
which  was  first introduced in \cite{G-RLRvUW} in the super-Poincar\'e case
and which is standard nowadays.}
 of weight $n $ is defined to be holomorphic 
on the north chart. It can be represented as 
\bea
\U^{(n)} (z, u) =  (u^{+\1})^n\, \U^{[n]} (z, \z) ~, \quad \qquad
\U^{ [n] } (z, \z) = \sum_{k=0}^{\infty} \U_k (z) \z^k~.
\label{arctic1}
\eea
The transformation law of $\U^{[n]}$ can be read off from 
eq. (\ref{harmult1}) by noting 
(see \cite{K,K2} for more details)
\bea
K^{ij}J_{ij}\, \U^{ [n] } (\z)
 =  \Big(  {K}^{++} (\z)\,\pa_\z + n\,K (\z) \Big) \U^{[n]}(\z)~,
\label{arctic2}
\eea
or equivalently
\bea
J_{\1 \1} \U_0 =0~, \qquad 
J_{\1 \1} \U_k &=&(k-1-n) \U_{k-1}~, \quad k>0 
\non \\
J_{\2 \2} \U_k &=&(k+1) \U_{k+1}~,  
\label{arctic3} \\
J_{\1 \2} \U_k &=&(\frac{n}{2}-k) \U_{k}~. \non 
\eea
It is important to emphasize  that the transformation law of  $\U^{[n]}$ 
preserves the functional structure of  $\U^{[n]}$  defined in (\ref{arctic1}).

The analyticity condition (\ref{ana2}) implies
\bea
\cD^{\2}_{\hat \a}  \U_0=0~, 
\qquad 
\cD^{\2}_{\hat \a}  \U_1 =\cD^{\1}_{\hat \a}  \U_0~.
\label{ana-arctic}
\eea
The integrability conditions for these constraints can be shown to be 
$J_{\1 \1} \U_0 = 0$ and $J_{\1 \1} \U_1 =- 2 J_{\1 \2} \U_0$, and they 
hold identically due to (\ref{arctic3}).
Using the algebra of covariant derivatives, eq. (\ref{covDev2spinor-}), 
one can deduce from (\ref{ana-arctic}) 
\bea
\Big( \cD^{\2}_{[\hat \a} \cD^{\2}_{\hat \b ]} +3{\rm i}\, \ve_{\hat \a \hat \b} \,
S^{\2 \2} \Big) \U_1 = 2{\rm i}\, \Big( \cD_{\hat \a \hat \b} - \frac{3}{2} n \,
\ve_{\hat \a \hat \b} \,
S^{\1 \2} \Big) \U_0~.
\eea

The smile-conjugate of $ \U^{(n)}$ is said to be 
an antarctic multiplet of weight $n $. It proves to be  holomorphic on the south
chart, while  in the north chart it has the form
\bea
\widetilde{\U}^{(n)} (z, u) &=&  (u^{+\2})^n\, \widetilde{\U}^{[n]} (z, \z)~, \qquad
\widetilde{\U}^{[n]} (z, \z) = \sum_{k=0}^{\infty} (-1)^k {\bar \U}_k (z)
\frac{1}{\z^k}~,
\label{antarctic1}
\eea
with $ {\bar \U}_k$ the complex conjugate of $U_k$.
Its  transformation follows from (\ref{harmult1}) by noting
\bea
 K^{ij}J_{ij}\,  \widetilde{\U}^{[n]} (\z)=  
 \frac{1}{\z^n}\Big(  {K}^{++} (\z) \,\pa_\z 
 + n\,K (\z) 
 \Big) \Big(\z^n\,\widetilde{\U}^{(n)} (\z)\Big)~.
\label{antarctic2}
\eea
The arctic multiplet $ \U^{[n]} $ and its smile-conjugate $\widetilde{\U}^{(n)} $ 
constitute a polar multiplet.

Our next example is a real $O(2n)$-multiplet, 
$\widetilde{H}^{(2n)} =H^{(2n)}$.
\bea
H^{(2n)}(z,u^+)&=&u^+_{i_1}\cdots u^+_{i_{2n}}\,H^{i_1\cdots i_{2n}}(z)
=\big({\rm i}\, u^{+\1} u^{+\2}\big)^n H^{[2n]}(z,\z) ~, 
\non \\
H^{[2n]}(z,\z) &=&
\sum_{k=-n}^{n} H_k (z) \z^k~,
\qquad  {\bar H}_k = (-1)^k H_{-k} ~. 
\label{o2n1}
\eea
The transformation  of $H^{[2n]} $ follows from (\ref{harmult1}) by noting
\bea
 K^{ij}J_{ij}\,  H^{[2n]} &=&  
 \frac{1}{\z^n}\Big(    {K}^{++} (\z) \,\pa_\z +2n\,K(\z)
 \Big) \Big(\z^n H^{[2n]} \Big)~.
\label{o2n2}
\eea
This can be seen to be equivalent to 
\bea
J_{\1 \1} H_{-n} =0~, \qquad 
J_{\1 \1} H_k &=&(k-1-n) H_{k-1}~, \quad  -n<k \leq n 
\non \\
J_{\2 \2} H_{n} =0~, \qquad
J_{\2 \2} H_k &=&(k+1+n) H_{k+1}~,  \quad  -n \leq k < n 
\label{o2n3} \\
J_{\1 \2} H_k &=& -k H_{k}~. \non 
\eea
The analyticity condition (\ref{ana2}) implies, in particular, the  constraints: 
$\cD^{\2}_{\hat \a}  H_{-n}=0$ 
and  $\cD^{\2}_{\hat \a}  H_{-n+1} =\cD^{\1}_{\hat \a}  H_{-n}$.
The corresponding integrability conditions can be shown to hold
due to (\ref{o2n3}).
The case $n=1$ corresponds to an off-shell tensor multiplet.

Our last example is a real tropical multiplet of weight $2n$:
\bea
U^{(2n)} (z,u^+) &=& 
\big({\rm i}\, u^{+\1} u^{+\2}\big)^n U^{[2n]}(z,\z) =
\big(u^{+\1}\big)^{2n} \big({\rm i}\, \z\big)^n U^{[2n]}(z,\z)~, \non \\
U^{[2n]}(z,\z) &=&
\sum_{k=-\infty}^{\infty} U_k (z) \z^k~,
\qquad  {\bar U}_k = (-1)^k U_{-k} ~. 
\label{trop-nj}
\eea
The SU(2)-transformation law of $U^{[2n]}(z,\z) $
 copies (\ref{o2n2}). To describe 
 a massless vector multiplet prepotential, one should choose $n=0$.
 Supersymmetric real Lagrangians correspond to the choice $n=1$, see below.

\sect{Supersymmetric action in  the Wess-Zumino gauge}

In our previous paper \cite{KT-M}, we formulated the supersymmetric action principle
in five-dimensional $\cN=1$ anti-de Sitter superspace ${\rm AdS}^{5|8}$. ${}$From  the supergravity point of view, the geometry of this  superspace is singled out by 
setting 
\bea
\cD^i_{\hat \a} S^{jk} =0~, \qquad F_{\hat a \hat b} = N_{\hat a \hat b} =0
\label{AdSlimit}
\eea
in the (anti)commutation relations
(\ref{covDev2spinor-}--\ref{covDev2spinor-3}), 
and the central charge decouples.\footnote{To make contact with the notation
used in \cite{KT-M}, one should represent $S^{ij}=  {\rm i}\,\o J^{ij}$.}
In a Wess-Zumino gauge, the action functional 
constructed in \cite{KT-M} is as follows:\footnote{We use the 
following definitions: $({\cD}^-)^4=-{1\over 96}\ve^{\hal\hbe\hga\hde}
\cD^-_{\hal}\cD^-_{\hbe}\cD^-_{\hga}\cD^-_{\hde}$
and $({\cD}^\pm)^2\,=\,\cD^{\pm\hal}\cD^\pm_{\hal}$.}
\bea
S&=&
-{1\over 2\pi} \oint {u_i^+\rd u^{+i}\over (u^+u^-)^4} 
\int \rd^5 x \,e \Big{[}~
(\cD^-)^4
-{25\over 24}\ri \,S^{--}(\cD^-)^2
+18S^{--}S^{--}
\Big{]}\cL^{++}\Big|~.~~~~~~~~~
\label{AdS}
\eea
Here $\cL^{++}$ is a real covariantly analytic superfield of weight $+2$, 
$\cD^-_{\hat \a} = u^-_i \cD^i_{\hat \a}$, $S^{--} = u^-_i u^-_j S^{ij}$, 
and the line integral is carried out over a closed contour in the space 
of $u^+$ variables. In the flat-superspace limit, $S^{ij} \to 0$, 
the action reduces to the 5D version  
of the projective-superspace action which was originally constructed in \cite{KLR}
and then reformulated in a projective-invariant form in \cite{S}.

As demonstrated in \cite{KT-M}, the action (\ref{AdS}) is uniquely fixed 
by either of the following two conditions: 
(i) supersymmetry; (ii) projective invariance.  
The latter means the invariance of $S$ under arbitrary
projective transformations of the form
\be
(u_i{}^-\,,\,u_i{}^+)~\to~(u_i{}^-\,,\, u_i{}^+ )\,R~,~~~~~~R\,=\,
\left(\begin{array}{cc}a~&0\\ b~&c~\end{array}\right)\,\in\,{\rm GL(2,\mathbb{C})}~.
\label{projectiveGaugeVar}
\ee
Thus, although the conditions (i) and (ii) seem to be unrelated at first sight, 
they actually appear to be equivalent. 
Below, we will put forward the principle of projective invariance in order 
to construct a supergravity extension of the above action. 

\subsection{Wess-Zumino gauge}

In supergravity theories, 
reduction from superfields to component fields is conveniently performed 
by choosing a  Wess-Zumino gauge \cite{WZ,WB}. 
Here we follow a streamlined procedure \cite{GGRS} of introducing  
the supergravity Wess-Zumino gauge \cite{WZ}
(originally given  in \cite{WZ} for the old minimal formulation of 
4D $\cN=1$ supergravity\footnote{For an 
alternative approach to impose  a supergravity Wess-Zumino gauge, 
see \cite{BK}.}).
The advantage of this approach is its universality
and independence from the dimension of space-time and 
the number of supersymmetries.

Given a superfield $U(z)=U(x,\q)$, it is standard to denote 
as $U|$ its  $\q$-independent component, 
$U|:= U(x,\q=0)$.
 The Wess-Zumino gauge 
for 5D $\cN=1$ supergravity  is defined by 
\bea
\cD_\ha|&=&\CD_\ha+\J_\ha{}^\hga_k(x) \cD^k_\hga|+\f_\ha{}^{kl}(x) J_{kl}+\cV_\ha(x) Z~,
\qquad
\cD^i_{\hat \a} |= \frac{\pa}{\pa \q^{\hat \a}_i}~.
\label{WZgauge1}
\eea
Here $\nabla_{\hat a}$ are  space-time covariant derivatives, 
\be 
\nabla_{\hat a} = e_{\hat{a}} + \o_{\hat{a}} ~, \qquad 
e_{\hat a} = e_{\hat a}{}^{\hat m} (x)\, \pa_{\hat m}~, 
\qquad  
 \o_{\hat{a}} =\hf \, \o_{\hat{a}}{}^{\hb \hc} (x) \,M_{\hb \hc}
 = \o_{\hat{a}}{}^{\hbe\hga} (x) \,M_{\hbe\hga}~,
\ee
with $e_{\hat a}{}^{\hat m} $ the component inverse vielbein, 
and $\o_{\hat{a}}{}^{\hb \hc}$ the Lorentz connection. 
Furthermore, $ \J_\ha{}^\hga_k$ is the component gravitino,
while $\f_\ha{}^{kl} = \F_\ha{}^{kl}|$ and $\cV_{\hat a} = V_{\hat a}|$ 
are the component SU(2) and central-charge gauge fields, respectively. 
The space-time covariant derivatives obey the commutation relations
\bea
\big[ \nabla_{\hat a} , \nabla_{\hat b} \big] = 
{\cT}_{\ha\hb}{}^{\hc} \, \nabla_{\hat c} 
+\hf \cR_{\hat{a}\hat{b}}{}^{\hat{c}\hat{d}}M_{\hat{c}\hat{d}}~. 
\eea
Here the space-time torsion can be shown to be 
\bea
{\cT}_{\ha\hb}{}^{\hc}&=&2\ri\,\ve^{jk}\,
\J_\ha{\,}^{\hga }_j(\G^\hc)_{\hga\hde}\J_\hb{\,}^\hde_k
~.
\eea
The latter occurs in the integration by parts rule:
\bea
\int \rd^5x\, e\,\CD_\ha U^\ha&=&\int \rd^5x\, e\,{\cT}_{\ha\hb}{}^{\hb}
\,U^\ha~, \qquad e^{-1} = \det \big(e_{\hat a}{}^{\hat m} \big)~.
\eea

Those supergravity gauge transformations (\ref{tau}) 
that survive in the Wess-Zumino gauge are described by the 
following equations: 
\bea
\cD^i_{\hat \a} K^{\hat \b}_j\big| &=& K^{\hat c}\big| \,T_{\hat c}{ \,}^i_{\hat \a  }{}^{\hat \b}_j \big|
+ \d^i_j \, K_{\hat \a}{}^{\hat \b}\big| +\d_{\hat \a}^{\hat \b} K^i{}_j \big|~, \qquad 
\cD^i_{\hat \a} K^{\hat b}\big| = 
-2{\rm i}\, (\G^{\hat b})_{\hat \a \hat \g} \,K^{\hat \g i}\big|~,
\non \\
\cD^i_{\hat \a} K^{\hat \b \hat \g}\big| &=& K^{\hat C}\big| \,
R_{\hat C}{ }^i_{\hat \a  }{}^{\hat \b \hat \g} \big|~, \qquad 
\cD^i_{\hat \a} K^{jk}\big| = K^{\hat C}\big| \,
R_{\hat C}{ }^i_{\hat \a  }{}^{jk} \big|~, \qquad 
\cD^i_{\hat \a} \t \big| =-2{\rm i}\, K^i_{\hat \a}\big|~.
\eea

\subsection{Action principle}
Let $\cL^{++}$ be a covariantly analytic real superfield of weight $+2$. 
We assume the existence (to be justified later on) of a locally supersymmetric 
and projective-invariant action associated with $\cL^{++}$.
Our main result is that the requirement of projective invariance uniquely determines
this action in the Wess-Zumino gauge, 
 provided the term of highest order in derivatives
is proportional to
\bea
 \oint {u_i^+\rd u^{+i}\over (u^+u^-)^4} 
\int \rd^5 x \,e \,
(\cD^-)^4\cL^{++}\Big|~.
\non
\eea
Direct and long calculations lead to the following 
projective-invariant action:
\bea
S(\cL^{++})&=&
-{1\over 2\pi} \oint {u_i^+\rd u^{+i}\over (u^+u^-)^4} 
\int \rd^5 x \,e \Bigg{[}
(\cD^-)^4
+ {\ri\over 4}\J^{\hal\hbe\hga-}\cD^-_\hga\cD^-_{\hal}\cD^-_\hbe
-{25\over 24}\ri \,S^{--}(\cD^-)^2
 \non\\
 &&~
-2(\S^{\ha\hb})_{\hbe}{}^\hga\J_{\ha}{}^{\hbe -}\J_{\hb}{}^{\hde-}
\cD_{{[}\hga}{}^{-}\cD_{\hde{]}}{}^{-}
-{\ri\over 4}\f^{\hal\hbe}{}^{--}\cD^-_{\hal}\cD^-_\hbe
+4(\S^{\ha\hb})^{\hal}{}_{\hga}\f_{{[}\ha}{}^{--}\J_{\hb{]}}{}^{\hga -}\cD^-_\hal
\non\\
&&~
-4\,\J^{\hal\hbe}{}_\hbe{}^{-}S^{--}\cD^{-}_\hal
+2\ri \,\ve^{\ha\hb\hc\hm\hn}(\S_{\hm\hn})_{\hal\hbe}
\J_{\ha}{}^{\hal -}\J_{\hb}{}^{\hbe-}\J_{\hc}{}^{\hga-}\cD_\hga^-
+18S^{--}S^{--}
\non\\
&&~
-6\ri \, \ve^{\ha\hb\hc\hm\hn}(\S_{\hm\hn})_{\hal\hbe}\J_{\ha}{}^{\hal -}
\J_{\hb}{}^{\hbe-}\f_{\hc}{}^{--}
+18\ri \, (\S^{\ha\hb})_{\hal\hbe}\J_{\ha}{}^{\hal-}\J_{\hb}{}^{\hbe -}S^{--}
\Bigg{]}\cL^{++}\Big|~.~~~~~~~~~
\label{Sfin}
\eea
The projective invariance of (\ref{Sfin}) is a result of miraculous cancellations.
The technical details will be given in \cite{KT-M2}.
In the AdS limit (\ref{AdSlimit}), the action reduces to (\ref{AdS}).

The important feature of our action  (\ref{Sfin}) is that it is practically ready for applications, 
that is, for a reduction from superfields to component fields.
It is instructive to compare (\ref{Sfin}) with the component actions in 4D $\cN=1$ supergravity
(see eq. (5.6.60) in \cite{GGRS} and eq. (5.8.50) in \cite{BK}).
It should be pointed out that one could also develop a harmonic-superspace 
formulation for 5D $\cN=1$ supergravity, in complete analogy with the 4D $\cN=2$ 
case \cite{GIOS}. But the action functional for supergravity-matter systems in 
harmonic superspace, as presented in  \cite{GIOS},  is given in terms of  the supergravity 
prepotential, and some work is still required to reduce it to a form 
useful for component reduction.

Without loss of generality, one can assume that the integration contour in 
(\ref{Sfin}) does not pass through the north pole $u^{+i} \sim (0,1)$.
Then, one can introduce the complex variable $\z$ as in (\ref{zeta}), 
and fix the projective invariance (\ref{projectiveGaugeVar}) as in 
(\ref{fix-u-}).
If we also represent the Lagrangian in the form 
\be
\cL^{++}(z,u^+)={\rm i} \,u^{+\1}u^{+\2}\cL(z,\z)
= {\rm i} (u^{+\1})^2 \,\z\,\cL(z,\z)~,
\label{L++-->L}
\ee
the line integral in (\ref{Sfin}) reduces to a complex contour integral of a function 
that is holomorphic almost everywhere in $\mathbb C$ except a few points.
Let us introduce several supergravity-matter systems. 

Given a set of $O(2)$ or tensor multiplets $H^{++I}  $,  with $I=1,\dots, n$, 
 their dynamics can be  generated by 
 a Lagrangian $\cL^{++}  = \cL \big(H^{++I} \big)$
that is a real homogeneous function 
of first degree in the variables $H^{++}$, 
\bea
H^{++I} \frac{\pa}{\pa H^{++I} } \,\cL \big(H^{++} \big) 
= \cL \big(H^{++}\big)~.
\eea
This is a generalization of superconformal tensor multiplets \cite{KLR,deWRV}.

Given a system of  arctic weight-one multiplets 
$\U^{+ } (z,u^+) $ and their smile-conjugates
$ \widetilde{\U}^{+}$, their dynamics can be   described by the Lagrangian
\bea
\cL^{++} = {\rm i} \, K(\U^+, \widetilde{\U}^+)~,
\label{conformal-sm}
\eea
with $K(\F^I, {\bar \F}^{\bar J}) $ a real analytic function
of $n$ complex variables $\F^I$, where $I=1,\dots, n$.
${}$For $\cL^{++}$ to be a weight-two real projective superfield, 
it is sufficient to  require 
\bea
 \F^I \frac{\pa}{\pa \F^I} K(\F, \bar \F) =  K( \F,   \bar \F)~.
 \label{Kkahler2}
 \eea
 This is a generalization of superconformal polar multiplets
\cite{K,KT-M,K2}.

Let $H^{++}$ be a tensor multiplet, and $\l$ an arctic weight-zero 
multiplet. Then, the action generated by $\cL^{++} = H^{++}\l$ 
vanishes, $S(H^{++}\l) =0$,
since the corresponding integrand in (\ref{Sfin})
can be seen to possess no poles (upon fixing the projective gauge). Thus
 \be
S\Big (H^{++}( \l +\tilde{\l})  \Big) =0~. 
\ee

A massless vector multiplet ${\mathbb V}(z,u^+)$ is described by a weight-zero tropical multiplet 
possessing the gauge invariance 
\be
\d {\mathbb V} =  \l +\tilde{\l}~, 
\ee
with $\l$ a weight-zero arctic multiplet. Given a tensor multiplet $H^{++}$, 
the Lagrangian $\cL^{++} = H^{++} \,{\mathbb V}$ generates a gauge-invariant action.

The minimal supergravity involves a vector multiplet associated with the central charge. 
If ${\mathbb V}(z,u^+)$ denotes the corresponding gauge prepotential, then 
the Lagrangian  $S^{++}\, {\mathbb V}$ leads to  gauge-invariant coupling.
If supersymmetric matter (including a compensator) is described by 
weight-one polar multiplets, then the supergravity-matter Lagrangian 
can be chosen to be
\be
\cL^{++}   = S^{++}\, {\mathbb V} +  {\rm i} \, K(\U^+, \widetilde{\U}^+)~,
\label{SUGRA-matter}
\ee
with the real function $K(\F, \bar \F) $
obeying 
the homogeneity condition (\ref{Kkahler2}).
Here we have used the fact that $S^{++}$ is covariantly analytic, as a consequence 
of (\ref{S-Bianchi2}).

As a generalization of the model given in \cite{KT-M}, 
a system of interacting arctic weight-zero multiplets 
${\bf \U}  $ and their smile-conjugates
$ \widetilde{ \bf{\U}}$ can be described by the Lagrangian 
\bea
\cL^{++} = S^{++}\,
{\bf K}({\bf \U}, \widetilde{\bf \U})~,
\eea
with ${\bf K}(\F^I, {\bar \F}^{\bar J}) $ a real function 
which is not required to obey any 
homogeneity condition. 
The action is invariant under K\"ahler transformations of the form
\be
{\bf K}({\bf \U}, \widetilde{\bf \U})~\to ~{\bf K}({\bf \U}, \widetilde{\bf \U})
+{\bf \L}({\bf \U}) +{\bar {\bf \L}} (\widetilde{\bf \U} )~,
\ee
with ${\bf \L}(\F^I)$ a holomorphic function.

In conclusion, we briefly discuss couplings of  vector multiplets to supergravity.
A U(1) vector multiplet can be described by its gauge-invariant field strength, $W(z)$,
which is a real scalar superfield obeying the  Bianchi identity 
(compare with the AdS case \cite{KT-M}),
\be
\cD^{(i}_{\hat \a} \cD_{\hat \b }^{j)}  W
= {1 \over 4} \ve_{\hat \a \hat \b} \,
\cD^{\hat \g (i} \cD_{\hat \g }^{j)}  W~. 
\label{Bianchi1}
\ee
The Bianchi identity implies that 
\be
G^{++} (z,u^+) = G^{ij}(z) \,u^+_iu^+_j:= 
{\rm i}\, \Big\{ \cD^{+ \hat \a} W \, \cD^+_{\hat \a} W 
+\hf  W \, 
( \cD^+)^2 W  +2{\rm i}\, S^{++} W^2 \Big\}
\label{YML}
\ee
is a composite real $O(2)$ multiplet, $\cD^+_{\hat \a} G^{++} =0$.
The coupling of the vector multiplet 
to supergravity is obtained by adding $G^{++}\, {\mathbb V} $
to the Lagrangian (\ref{SUGRA-matter}). For the central-charge vector multiplet, 
its $O(2)$-descendant ${\mathbb G}^{++}$ reduces to $(-2)S^{++}$, as a result 
of a super-Weyl gauge fixing ${\mathbb W}=1$  implicitly made
in Howe's formulation \cite{Howe5Dsugra}.

\sect{$\bm{\L}$-group and supersymmetric action}

In this section, we formulate a locally supersymmetric action 
underlying the dynamics of supergravity-matter systems.
Our construction  has some analogies with  
the prepotential formulation for 4D $\cN=1$ supergravity 
reviewed in \cite{GGRS,BK}, specifically the supergravity  gauge or $\L$ group
\cite{OS,SG}.
It is also analogous to  the harmonic-superspace 
approach to  the minimal multiplet for 4D $\cN=2$ supergravity \cite{GIOS}
as re-formulated in appendix A of  \cite{KT}.

\subsection{$\bm{\L}$-group}
Consider the space of analytic multiplets of weight $n$, $\hat{Q}^{(n)}(z,u^+)$, 
in flat superspace ${\mathbb R}^{5|8}$.
Such superfields are defined by the equations (\ref{ana}--\ref{harmult1})
in which the curved-superspace covariant derivatives $\cD^+_{\hat \a}$ 
have to be replaced by flat ones,  $D^+_{\hat \a}$. Introduce a Lie algebra of 
first-order differential operators  acting on this linear functional space. 
Such an operator $\L$  generates an infinitesimal variation  of $\hat{Q}^{(n)}(z,u^+)$
of the form:\footnote{For simplicity, the analytic multiplets are chosen to be
independent of the central charge, $Z\hat{Q}^{(n)}=0$. It is not difficult to extend 
out analysis to the general case; compare also with \cite{KT}.}
\bea
\d \hat{Q}^{(n)} = \L \hat{Q}^{(n)}~,
\qquad 
\L = \L^{\hat m} \pa_{\hat m}
- \frac{1}{(u^+u^-)} \Big(\L^{+\hat \m} D^-_{\hat \m} +\L^{++}D^{--}\Big)
+n\,\S~.
\label{lambda-group}
\eea
The transformation parameters $ \L^{\hat m}$, $\L^{\hat \m}  $,
$\L^{++}$ and $\S$ are  such that the  variation of  $\hat{Q}^{(n)}$, 
$\L \hat{Q}^{(n)}$, is also a flat analytic  superfield of weight $n$.
The requirement of analyticity, $D^+_{\hat \a} \hat{Q}^{(n)}=0$,
can be seen to imply 
\bea 
D^+_{\hat \m } \L^{\hat \n \hat  \g} &=&
8{\rm i} \Big( \d_{\hat \m}^{[ \hat \n} \,\L^{+\hat \g ]}
+\frac{1}{4} \ve^{\hat \n \hat \g}\, \L^{+\hat \m} \Big)~,
\quad D^+_{\hat \m } \L^{\hat \n } =  \d_{\hat \m}^{\hat \n} \,\L^{++}~,
\quad D^+_{\hat \m } \L^{++ } = D^+_{\hat \m } \S =0~.~~~~~~~
\eea
The requirement of $\L \hat{Q}^{(n)}$ to be independent of $u^-_i$ can be shown 
to hold if
\bea
\frac{\pa}{\pa u^{-i} } \L^{\hat m} = \frac{\pa}{\pa u^{-i} } \L^{+\hat \m} =
\frac{\pa}{\pa u^{-i} } \L^{++}=0~,
\quad 
u^{-i}\frac{\pa}{\pa u^{-i} } \S = 0~, \quad D^{++}\S = \frac{\L^{++} }{(u^+u^-) } ~.
~~~
\eea
It is also clear that the requirement of $\L \hat{Q}^{(n)}$ to have weight $n$ 
holds if $\L^{\hat m} $, $\S$, $ \L^{+\hat \m} $ and $ \L^{++}$ are homogeneous 
functions of $u^+$ of degrees $0, 0, 1$ and 2 respectively.

A  solution to the above constraints is:
\begin{subequations}
\bea
\L^{\hat \m \hat  \n} &=& {\rm i}\, 
\Big( D^{+\hat \m }D^{+\n} +\frac{1}{4} \, \ve^{\hat \m \hat \n} 
(D^+)^2 \Big)\O^{--}~, \qquad
\L^{+\hat \m } = -\hf D^{+\hat \m}(D^+)^2 \O^{--}~,\\
\L^{++ } &=& -\frac{1}{8} (D^+)^2(D^+)^2 \O^{--}~,\qquad \qquad 
\eea
\end{subequations}
as well as
\bea
2\S = \pa_{\hat m}\L^{\hat m}
+\frac{1}{(u^+u^-)} \Big(D^-_{\hat \m}\L^{+\m} - D^{--}\L^{++}\Big)~.
\eea
Here the parameter  $\O^{--}$ is required to be  
(i) independent of $u^-$; and (ii) homogeneous in 
$u^+$ of degree $-2$.

Let $\hat{\cL}^{++} (z,u^+)$ be 
a real  analytic superfield of weight $+2$. 
Its transformation can be seen  to be a total derivative:
\bea
\L \, \hat{\cL}^{++}&=& \pa_{\hat m}\big(\L^{\hat m}\hat{\cL}^{++}\big)
+\frac{1}{(u^+u^-)} D^-_{\hat \m}\big( \L^{+\m} \hat{\cL}^{++}\big)
-\frac{1}{(u^+u^-)}  D^{--} \big( \L^{++}\hat{\cL}^{++}\big)~.
\label{Lag-L-tra}
\eea
Therefore, the   following functional 
\bea
S=-\frac{1}{ 2\pi}
\oint {u_i^+\rd u^{+i}\over (u^+u^-)^4}\int\rd^5 x\,({D}^-)^4\hat{\cL}^{++}
\label{flataction}
\eea
is invariant under (\ref{Lag-L-tra}).

A crucial element in the above construction is the Lie algebra 
of first-order operators of the form 
\be
\cA = (D^{++} \S) \,D^{--} - n\,\S, \quad 
u^{+i}\frac{\pa}{\pa u^{+i} } \S=u^{-i}\frac{\pa}{\pa u^{-i} } \S=0~,
\quad (D^{++})^2 \S=0~,
\ee
which act on the space $\cF^{(n)}$ of functions $\cQ^{(n)}(u^+)$ 
being homogeneous in $u^+$ of order $n$.
This algebra can be viewed to be a  gauging of 
the algebra ${\rm su}(2) \oplus {\mathbb R}$ naturally acting on $\cF^{(n)}$, 
where $\mathbb R$ corresponds to infinitesimal scale transformations.

\subsection{A partial solution to the constraints}
Given a covariantly analytic superfield $Q^{(n)}$, 
eq. (\ref{analyt}) implies that the covariant derivatives $\cD^+_{\hat \a} $
and $Q^{(n)}$ can be represented as follows: 
\begin{subequations}
\bea
\cD^+_{\hat \a} &=& {\rm e}^{\cH} \,\D_{\hat \a}^+ 
{\rm e}^{-\cH}  ~, \qquad 
\D_{\hat \a}^+ = N_{\hat \a}{}^{\hat \b}  D^+_{\hat \b} 
+\hat{\O}^+_{\hat \a} {\,}^{\hat \b \hat \g} M_{\hat \b \hat \g}
+ \hat{ \F}^{-}_{\hat \a} J^{++}
~, \\
Q^{(n)} &=& {\rm e}^{\cH} \,\hat{Q}^{(n)}~,
\qquad D^+_{\hat \a} \hat{Q}^{(n)}=0~.
\eea
\end{subequations}
Here $\cH (z,u)$ is some first-order differential operator, 
and  $\hat{Q}^{(n)} (z,u^+)$  a flat analytic superfield of weight $n$.
The covariant derivatives are left invariant under 
 gauge transformations of the prepotentials of the form:
 \bea
 \d {\rm e}^{\cH} = -{\rm e}^{\cH} \, {\bm \L}~,
 \qquad \d \D_{\hat \a}^+  = [ {\bm \L}, \D_{\hat \a}^+ ]~, \qquad 
 {\bm \L} = \L + \r^{-\hat \m} D^+_{\hat \m} 
+\r^{\hat \b \hat \g} 
M_{\hat \b \hat \g} + \r^{--}J^{++}~,~~~~~~
 \eea
where
$\L$ is the same as in eq. (\ref{lambda-group}), 
while the parameters $\r^{-\hat \m}$, $\r^{\hat \b \hat \g} $ and $\r^{--}$
are arbitrary modulo homogeneity conditions. 
The analytic superfield $\hat{Q}^{(n)}$ transforms as in (\ref{lambda-group}).
The variation $\hat{Q}^{(n)}$ involves not only general coordinate 
and local ${\rm SO}(4,1)\times {\rm SU}(2)$ transformations, 
but also Weyl transformations.

Let $\cL^{++}$ be the covariantly analytic Lagrangian in
the  action  (\ref{Sfin}). It can be represented  in the form 
$\cL^{++} = {\rm e}^{\cH} \,\hat{\cL}^{++}$, for some 
flat analytic superfield $\hat{\cL}^{++}(z,u^+)$ of weight $+2$.
Then, the action (\ref{flataction}) generated by  $\hat{\cL}^{++}$
is locally supersymmetric and projective-invariant. 
In the Wess-Zumino gauge, it should turn into  (\ref{Sfin}).

In order to extend our construction to the case of 4D $\cN=2$ supergravity, 
one should build on the superspace formulation for 
the minimal supergravity multiplet given in \cite{Howe}.

\noindent
{\bf Acknowledgements:}\\
This work is supported  in part
by the Australian Research Council and by a UWA research grant.

\small{

}

\end{document}